\documentclass[runningheads]{llncs}
\usepackage[T1]{fontenc}
\usepackage{graphicx}
\usepackage{float}
\usepackage{comment}
\usepackage{pgfplots}
\pgfplotsset{compat=1.17}

\begin{document}
\title{Quantum Machine Learning in Precision Medicine and Drug Discovery - A Game Changer for Tailored Treatments?} %Opportunities and Challenges of Quantum Computing and Quantum Machine Learning in Precision Medicine
\titlerunning{Quantum Computing in Precision Medicine}
\author{Markus Bertl\inst{1,2,3}\orcidID{0000-0003-0644-8095} \and
Alan Mott\inst{1} \and
Salvatore Sinno\inst{1,4}\orcidID{0009-0002-9177-5161} \and
Bhavika Bhalgamiya\inst{1}\orcidID{0009-0001-8586-4531}
}
\authorrunning{M. Bertl et al.}

\institute{NextGen Computing Research Group, Unisys, Blue Bell, Pennsylvania, USA %\email{\{markus.bertl,alan.mott,salvatore.sinno,shruthi.thuravakkath@unisys.com\}@unisys.com}
\and
Department of Health Technologies, Tallinn University of Technology, Estonia
\and
Department of Information Systems and Operations Management, Vienna University of Economics and Business, Austria
\and
School of Computing, Newcastle University, UK
}
\maketitle              % typeset the header of the contribution
\begin{abstract}
%The abstract should briefly summarize the contents of the paper in 150--250 words.
The digitization of healthcare presents numerous challenges, including the complexity of biological systems, vast data generation, and the need for personalized treatment plans. Traditional computational methods often fall short, leading to delayed and sometimes ineffective diagnoses and treatments. Quantum Computing (QC) and Quantum Machine Learning (QML) offer transformative advancements with the potential to revolutionize medicine by leveraging quantum mechanics principles. This paper summarizes areas where QC promises unprecedented computational power, enabling faster, more accurate diagnostics, personalized treatments, and enhanced drug discovery processes. However, integrating quantum technologies into precision medicine also presents challenges, including errors in algorithms and high costs. We show that mathematically-based techniques for specifying, developing, and verifying software (formal methods) can enhance the reliability and correctness of QC. By providing a rigorous mathematical framework, formal methods help to specify, develop, and verify systems with high precision. In genomic data analysis, formal specification languages can precisely (1) define the behavior and properties of quantum algorithms designed to identify genetic markers associated with diseases. Model checking tools can systematically explore all possible states of the algorithm to (2) ensure it behaves correctly under all conditions, while theorem proving techniques provide mathematical (3) proof that the algorithm meets its specified properties, ensuring accuracy and reliability. Additionally, formal optimization techniques can (4) enhance the efficiency and performance of quantum algorithms by reducing resource usage, such as the number of qubits and gate operations. Therefore, we posit that formal methods can significantly contribute to enabling QC to realize its full potential as a game changer in precision medicine.

\end{abstract}
%\tableofcontents
%
%
%
\section{Introduction}
\label{sec:Intro}
The healthcare digitization faces numerous challenges, including the complexity of biological systems, the vast amount and quality of data generated \cite{bertl2023evaluation}, and the need for personalized treatment plans \cite{goetz2018personalized}. Traditional computational methods often struggle to process and analyze data efficiently \cite{bote2019deep}, leading to delays in diagnosis and treatment. Quantum computing (QC) and quantum machine learning (QML) represent transformative advancements with the potential to revolutionize precision medicine \cite{baiardi2023quantum}. QC promise unprecedented computational power through the use of quantum mechanic principles like superposition, entanglement and quantum tunneling \cite{bernhardt2019quantum}. Based on quantum mechanics, novel algorithms can be developed to tackle complex biological data, leading to more accurate diagnostics, personalized treatments, and enhanced drug discovery processes \cite{chow2024quantum}. However, the integration of QC into precision medicine also presents significant challenges, including technical limitations, high costs, and the need for interdisciplinary collaboration. This paper explores the opportunities and challenges associated with leveraging QC and QML to advance precision medicine, aiming to provide a comprehensive overview of the current landscape and future directions.\par
%Why Precision and personalized medicine?
The healthcare industry is undergoing a paradigm shift from a one-size-fits-all approach to a more tailored and individualized model of care \cite{bertl2023future,bertl2023challenges}. P5 medicine (predictive, preventive, personalized, participatory and psycho-cognitive medicine) \cite{gorini2011p5} is at the forefront of this transformation, aiming to provide treatments that are specifically designed for individual patients based on their unique genetic, environmental, and lifestyle factors \cite{roden2013genomic}. 
Traditional medical practices often rely on generalized treatment protocols that do not account for the significant variability among patients. This approach can lead to suboptimal outcomes, as treatments that are effective for one group of patients may be ineffective or even harmful for others \cite{razzak2020big}. This lack of personalization in treatment plans can result in prolonged illness, increased healthcare costs, and a higher incidence of adverse drug reactions. 
This is, in large part, because genetic variability plays a crucial role in how individuals respond to medications and treatments \cite{madian2012relating}. For example, certain genetic mutations can affect drug metabolism, leading to variations in drug efficacy and safety.
P5 medicine can leverage genetic information to identify these variations and tailor treatments accordingly. By understanding a patient’s genetic makeup, healthcare providers can predict which treatments will be most effective and avoid those that may cause harm. Because of the aging and growing populations, diseases and health care costs rise, making disease prevention even more important \cite{mendelson1993effects,de2013effect}.
Many complex diseases, such as cancer, cardiovascular diseases, and neurological disorders, have multifactorial causes that involve intricate interactions between genes, environment, and lifestyle. Traditional approaches often fall short in addressing these complexities. P5 medicine, through advanced technologies such as genomic sequencing and methods of bioinformatics, can unravel these interactions and provide insights into the underlying mechanisms of diseases. This knowledge enables the development of targeted therapies that address the root causes of diseases rather than just alleviating symptoms.
Personalized medicine also has the potential to advance preventive care. By identifying individuals at high risk for certain diseases based on their genetic and environmental profiles, healthcare providers can implement early interventions to prevent the onset of diseases. This proactive approach can significantly reduce the burden of chronic diseases and improve overall population health.\par
The ultimate goal of P5 medicine is to improve patient outcomes. By tailoring treatments to the individual characteristics of each patient, healthcare providers can achieve higher treatment success rates, reduce the incidence of adverse effects, and enhance the overall quality of care. Personalized medicine empowers patients by involving them in their own care decisions and providing them with treatments that are specifically designed for their needs.
%Problem Statement
Apart from ethical, legal and social issues, as well as regulatory and policy challenges, data integration and management as well as associated costs are a crucial barrier for the large-scale implementation of P5 medicine. The vast amount of data, such as genomic and phenotypic data, chemical/pharmaceutical data and data from electronic health records (EHR), require advanced computational tools for evaluation which are often associated with high hardware costs. 

%TBD: Classical Computing's Limitations and how Quantum can overcome them 
\subsection{Limitations of a Classical Approach to Solving Complex Problems}
The limitations of using classical techniques to solve complex problems are not just confined to the limitations of classical computing hardware. Classical mathematical algorithms themselves also have their own limitations. We will briefly visit both these areas in this section.

\subsubsection{Limitations of Classical Hardware}
The reader may very well be familiar with \textit{Moore's Law}, which states that the number of transistors on a microchip doubles about every two years, though the cost of computing is halved. There is, however, a physical limitation to this. Transistors cannot reduce in size beyond a few atoms. Microchip technology has already progressed to the point where this size limitation may already have been reached and Moore's Law may have already plateaued. Moore's Law is tied to \textit{Dennard Scaling}, which states that as the dimensions of a device go down, so does power consumption. However, this again has physical limits. With smaller and smaller devices, the chances of current leakage across the device becomes ever more likely and increases the chance of thermal runaway (and hence the destruction) of the device through overheating. In addition, classical computer architecture still largely adheres to the \textit{Von Neumann Architecture}, where the microprocessor and memory are connected via a narrow bus, creating what is known as the \textit{Von Neumann Bottleneck}: As demand for faster computing power to solve ever more complex problems increases, the demand on this bus to exchange data between the microprocessor and its memory exceeds its capability. One approach in resolving this is to use a highly parallelized and distributed architecture where multiple processors (Graphical Processing Units or GPUs) execute commands and process data in parallel. One obvious disadvantage of this is cost. At the time of writing, GPUs are approximately three times the price of CPUs. Moreover, GPUs are designed to perform one particular task at scale (render graphics) and are not optimized for general computing tasks. The high processing demands made on GPUs also increases their thermal power consumption, increasing the system's overall energy costs. In addition, despite their parallel architecture with many Processing Elements (PEs) working in parallel inside the GPU, they still suffer from limitations imposed by memory bus bandwidth issues within the GPU itself. These architectures help, and improve perfomance in certain contexts, namely, where a task can be broken down into subtasks that can be executed in parallel and the completion of one task is not dependent on the completion of another subtask. But they are not a generalised solution to the limitations of the Von Neumann architecture.

The rising demand for faster computing power to solve increasingly complex problems in a reasonable time frame is pushing the capabilities of classical computing hardware to its physical limits. 

\subsubsection{Limitations of Classical Algorithms}
Hardware limitations are not the only issues affecting our abilities to solve problems of ever increasing complexity. Our current methods of solving such problems also suffer from limitations. In other words, the technique (or algorithm) used to solve a problem in and of itself introduces limitations. An example of this is \textit{Time Complexity}. We discuss this here in overview only in order to outline how new algorithms, previously unavailable top us as they rely on having access to quantum states, can provide us with speed increases in finding problem solutions.

Many problems that require a mathematical solution have a defined rate of increase in time needed to solve that problem as that problem's complexity increases. For example, factoring a number is a problem who's complexity, and hence the time needed to solve the problem, increases as the number to be factored gets larger and larger. This issue is known as \emph{Time Complexity}. The rate of increase in the time to solve a problem as its complexity increases is usually categorised as either requiring \emph{polynomial} or \emph{exponential} time. Exponential time increases in accordance with

$$f(x)=b^x, \textit{ for some } b,$$ 

\noindent while polynomial time increases in accordance with 

$$f(x)=x^k, \textit{ for some } k.$$

Fig. \ref{fig:Time Complexity} below shows how much more aggressively the time taken to solve more complex problems increases. With classical algorithms, we may alleviate this time penalty through the use of increased classical compute resources (e.g. faster processors, more memory etc.). Such an approach could, for example, change the time scale for solving a problem from seconds to milliseconds. But, as stated above, we are reaching the limit of classical computing hardware capability and yet we continue to demand solutions to ever more complex problems.
\begin{figure}[H]
\begin{tikzpicture}
    \begin{axis}[
        xlabel={complexity},
        ylabel={time},
        xmin=0, xmax=27,
        ymin=0, ymax=135, % Adjusted to show the range of the functions
        axis lines=middle,
%        legend pos=north west,
        xtick={0,5,...,20},
        ytick={0,10,...,120},
        title={Time Complexity}
    ]
    \addplot[
        domain=0:7, 
        samples=100, 
        color=blue,
    ]
    {2^x};
    \addlegendentry{Classical}

    \addplot[
        domain=0:14, 
        samples=100, 
        color=red,
    ]
    {2^(x/2)};
    \addlegendentry{Quantum}
    \end{axis}
\end{tikzpicture}
    \centering
    \caption{Time Complexity. With classical algorithms, the time taken to solve a problem may increase in superpolyniomial time (i.e. exponentially) with $2^{n}$. With quantum algorithms, the time may only increase, as we explain in \ref{sec:grover} based on the example of Grover's algorithm, with $2^{n/2}$. }
    \label{fig:Time Complexity}
\end{figure}
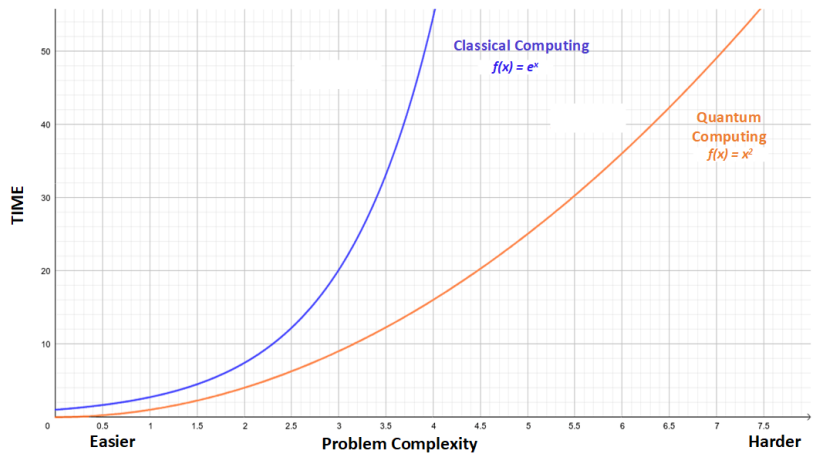
For some problems, we need to develop new algorithms in order to solve very complex examples of them in a reasonable time frame and overcome the issues of time complexity.

\subsubsection{Grover's Algorithm - An Example of Significantly Decreasing Time Complexity}\label{sec:grover}
As we have seen, solving very complex problems requires a solution to the time complexity issue. Due to us hitting the limits of Moore's law and the like, masking this issue by investing more money in faster and more performant classical hardware is now a law of diminishing returns. Quantum algorithms, however, can in some instances decrease a problem's time complexity. To illustrate this, we will briefly look at Grover's Algorithm. This is a quantum algorithm that can reduce the time complexity of searching through data.
While Grover himself said in his paper that this algorithm does not solve search problems in polynomial rather than superpolynomial time, it does, however, reduce the time taken to search unstructured data significantly over classical techniques \cite{grover1996fast}. We reference Grover's algorithm here as it is easy to grasp the speed increase this algorithm provides, namely $O(\sqrt{2^n})$ or in another format $O(2^{n/2})$ as opposed to $O(2^n)$ for some number of search items $n$. Although this does not reduce the complexity from exponential to polynomial, it provides a practically important speed up of quantum compared to classical computing.

\begin{figure}[H]
    \centering
    \includegraphics[width=1\linewidth]{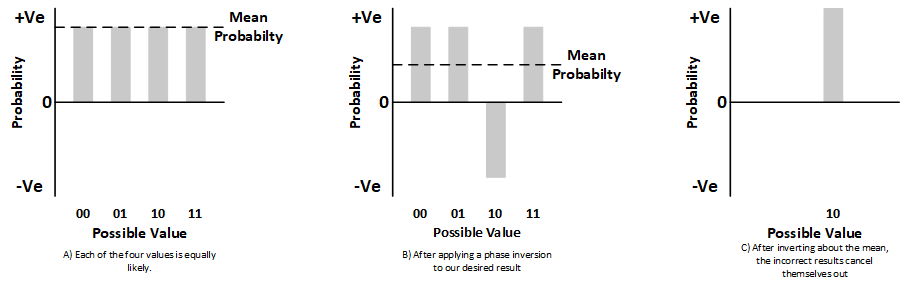}
    \caption{Grover's Algorithm}
    \label{fig:grover}
\end{figure}

Consider the problem where we have \textit{N} number of data items (\textit{n}), and we need to find one specific item in this set (we'll call it \textit{W}). Using classical techniques, we would need to examine each item \textit{n} to test whether it is \textit{W}. We use \textit{t} to denote the time it takes to test a data item \textit{n} to ascertain whether it equals \textit{W} or not . The total time \textit{T} is the time it takes us to find \textit{W} in a list of data items \textit{n} in a set of size \textit{N}. Thus, our best case scenario is given by:\[ T(W) = t \] where we are lucky enough that the first data item we test happens to be \textit{W}. However, our worst-case scenario is:\[ T (W) = t\\ N \] where the item we are searching for is the last item in set \textit{N}. The average time it will take us to find \textit{W} is:\[T(W) = tN/2\] Using this classical search technique, we cannot improve on this. However, Grover's algorithm exploits the quantum phenomena of superposition to optimise our search (see Fig. \ref{fig:grover}).
\begin{remark}
Superposition describes the quantum state where a property of a quantum particle (e.g. an electron or photon) is not in one specific state, but rather exists in a state of probabilities that the property is one value or another. In QC, such quantum particles that we can manipulate in this way are referred to as \textit{qubits}.
\end{remark}
Let's suppose we are searching for a specific number between 0 and 3. We will denote these four values in binary, i.e. 00, 01, 10 and 11. That gives a set of possible choices of \textit{N=4}. Furthermore, let us suppose that we are searching for the number 2 (10 in binary). We can use four quantum particles (i.e. qubits) to each represent the four numbers we need to search through. At a very high level, and avoiding all the advanced mathematics and quantum theory behind it, the process is as follows:
\begin{enumerate}
    \item We put each qubit in a state of superposition, whereby the probability of that qubit being any of our 4 numbers is equal
    \item We then apply a function that inverts our required answer (in this case 2 or 10)
    \item We then invert the probabilities about the mean value. The three incorrect answers cancel themselves out, and we are left with our desired result
\end{enumerate}
In essence, Grover's algorithm allows us to search all of our items \textit{n} in the set with \textit{N} items at once, rather than examining each individually. As Groover explains in his paper \cite{grover1996fast}, this gives us an average search time of \[T(W)=t\ \sqrt{N} \] as opposed to \[T(W)=t N/2\] So, if we had 100 items to search in order to find \emph{}{W}, using classical techniques we would take on average the time it takes to search 50 items to find it. Using Grover's algorithm, it only takes the time to search 10.\\
\\
The key takeaway here is that by exploiting quantum states such as superposition and entanglement, new algorithms may be developed that provide speed increases that classical solutions cannot achieve. The media makes much of how "fast" quantum computers are (or will be). But the reality, as Grover's algorithm demonstrates, is that it is actually the quantum algorithm which is needed for the performance improvements claimed.\\
\\
Quantum algorithms therefore provide us with the advanced tools we need to analyse and evaluate vast quantities of data, as well as solve highly complex problems in reasonable time frames. However, we cannot solve the equations associated with quantum algorithms using classical hardware. Classical hardware does not have access to fundamental particles and their quantum states. We therefore need quantum computers in order to execute these algorithms and achieve the quantum advantage we require to effectively deliver the improved health outcomes promised by P5 medicine.

%protein folding is even not possible with classical
%Applications based on machine-\cite{bertl2021predicting,bertl2023finding} and deep learning \cite{bertl2024evaluation} already showed promising results in health research. However, when those algorithms are applied to tasks in predictive or preventive medicine, \cite{razzak2020big} has found that they often struggle to perform well and/or are computational ineffective or even infeasible. Research suggests that Quantum Computing and Quantum Machine learning can overcome those challenges. \cite{cerezo2022challenges} \cite{houssein2022machine} \cite{flother2023state} \cite{ur2023quantum} \cite{solenov2018potential}

\subsection{Quantum Computers}
A Quantum Computer is a device capable of performing quantum calculations. Such calculations include Grover's Algorithm discussed earlier. Shor's Algorithm, which provides computational speed up for factoring numbers, is another. As we have seen, quantum algorithms can reduce time complexity and allow some problems to be solved quicker. However, the computation of such algorithms must be carried out on devices that allow the manipulation of the quantum properties of fundamental particles such as electrons and photons. Such manipulation involves, for example, putting them in superposition. The device that allows us to do this the quantum computer.\\
Quantum computers utilise fundamental sub-atomic particles such as electrons and photons and manipulate their quantum properties in order to process information. Whereas the basic unit of information in a classical computer is a \textit{bit}, with quantum computers these are known as \textit{qubits}.
At present, there are two families of quantum computer; gate-based quantum computers and quantum annealers.

\subsubsection{Gate-based Quantum Computing}
Gate-based quantum computers work by applying a series of functions known as \textit{gates} to its qubits. Each gate manipulates the state of the qubit. For example, the Hadamard gate puts a qubit into superposition, where as two qubit gates, like the Controlled-NOT (CNOT) gate, entangles two qubits. A collection of gates manipulating a collection of qubits in this way is called a \textit{quantum circuit}. Quantum gates and circuits are designed to implement the mathematical functions that comprise quantum algorithms. These are then in turn implemented on the quantum computer using software programming libraries such as Qiskit, Q\#, and Cirq.

\subsubsection{Quantum Annealing}
Quantum Annealers are a type of quantum computers which uses quantum annealing phenomena where the existing energy state of any quantum mechanical system can be manipulated by external magnetic fluctuations. The quantum mechanical system of qubits is initialised into an energy state that represents the problem to be solved. Physical systems can be described by a formula called the Hamiltonian, that describes the total mechanical energy (both kinetic and potential) within that system. Each individual system is described with its own version of the Hamiltonian equation. In the case of the quantum annealer, the Hamiltonian of the initial state represents the problem to be solved, encoded in such a way that the system's minimum energy level (or \textit{Ground State}) represents the solution. From such a starting position, the system will have only a finite set of energy states that it can be in. This finite set of energy levels is known as the system's energy landscape. The quantum mechanical system is then allowed to "evolve", exploring this energy landscape until the global minimum is achieved for the given problem Hamiltonian. The state of the system is then read out as the solution to the problem.

\subsubsection{Quantum Gate vs Quantum Annealing}
Quantum Annealing is, compared to gate-based QC, not Turing-complete. At present, gate-based quantum computers must be considered a nascent technology. It has been estimated that in order to crack RSA 2048 encryption, a quantum computer with 4099 qubits would be required. As of 2024, the largest gate-based quantum computer has 1121 qubits (IBM's Heron). D-Wave, however, has quantum annealers containing in excess of almost 5670 qubits  on Advantage Machines or their current generation of quantum annealers. However, due to the nature of how quantum annealers work, they are not considered a viable platform for universally solving problems of all classes. They are primarily useful for solving the combinatorial optimisation class of problems (those where there are many viable solutions, but only one is considered "best", or optimal). Whilst their application is limited, quantum annealers are, however, a commercially viable platform for solving such problems at scale.

\subsubsection{Quantum Machine Learning}
\begin{figure}[H]
    \centering
    \includegraphics[width=0.39\linewidth]{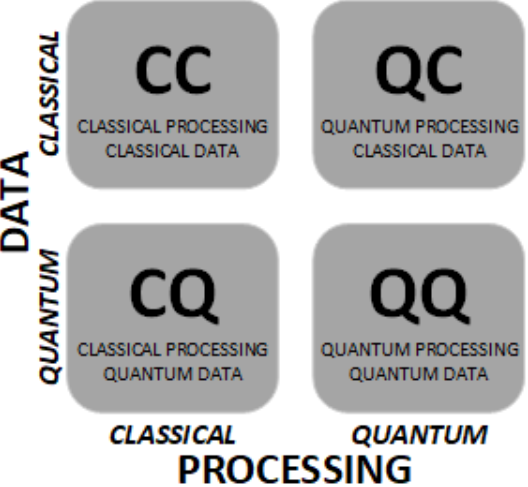}
    \caption{The four subsets of QML.}
    \label{fig:qml_types}
\end{figure}

\begin{figure}[h]
    \centering
    \includegraphics[width=0.2\linewidth]{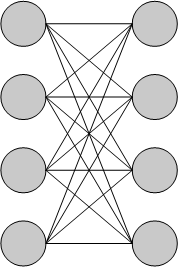}
    \caption{Single Cell contained within the Chimera Topology in D-Wave's second generation Quantum Annealer}
    \label{fig:Chimera}
\end{figure}

QML, as its name suggests, involves combing ML algorithms with QC. Machine Learning (ML) uses algorithms to analyse data from past events to predict future outcomes. ML can involve the analysis of vast amounts of data and this can be a challenge for classical computers. Moreover, ML is generally more accurate in its predictions the more data it is given to learn from \cite{bertl2024evaluation}. QML therefore represents a field of technology where quantum computers can be leveraged to allow machine learning systems to process massive amounts of data and hence produce more accurate results in a reasonable time frame. Four types, or subsets, of QML are identified, depending on how QC and ML are combined (see Fig. \ref{fig:qml_types}):
\begin{itemize}
    \item[\textbullet] \textbf{CC - Classical Processing, Classical Data}
Utilising classical algorithms and compute power to process data generated by classical means. This is the predominant and classical form of ML.
    \item[\textbullet] \textbf{QC - Quantum Processing, Classical Data}
Using quantum computing and algorithms to process data from classical sources. Otherwise known as \textit{quantum-enhanced} ML, this is the main area of current development in applying quantum techniques to ML.
    \item[\textbullet] \textbf{CQ - Classical Processing, Quantum Data}
This involves using classical ML techniques to learn from quantum states. Classifying quantum states output by a quantum system using classical hardware and classical algorithms is an example of this.
    \item[\textbullet] \textbf{QQ - Quantum Processing, Quantum Data}
Using quantum computers and algorithms to analyse and learn from data generated by a quantum system.
\end{itemize}
Quantum equivalents of classical ML algorithms have also been developed. For example, the Support Vector Machine (SVM) is an algorithm used widely in supervised ML. Its quantum equivalent, the Quantum Support Vector Machine (QSVM), can leverage Grover's algorithm to search all possible solutions to find the dividing lines (or, more formally, the hyperplanes) that classify the supplied input data points into sets (in other words, unique classes or output labels). Embedding Grover's algorithm in this way can provide a quadratic speed up to the SVM algorithm \cite{anguita2003quantum}.\\
Additionally, the very architecture of Quantum Processing Units (QPUs) lends itself to certain machine learning techniques. The Artificial Neural Network (ANN) is an ML technique that mimics the way the brain learns and is based on artificial representations of neurons and synapses. An example architecture of a particular type of ANN, the Feed Forward Neural Network is shown in Fig. \ref{fig:ANN_Architecture} below. This example of a ANN is comprised of three layers, the input, hidden and output layers. Each node in each layer is connected to all the nodes in the next layer, but not to each other. Each node applies a simple function to its inputs before triggering the nodes in the next layer. When learning, the output is compared to the known input. If different, the values associated with each node's function are tweaked and the process repeated. This continues until the output matches the input and the system can be said to have learned the input.
\begin{figure}[h]
    \centering
    \includegraphics[width=0.6\linewidth]{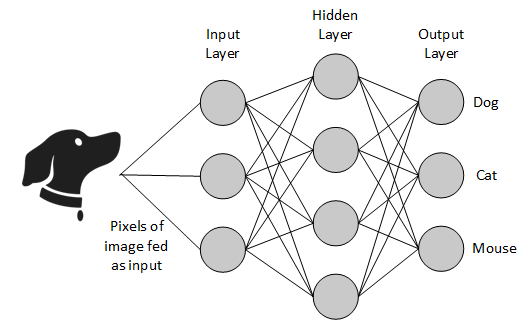}
    \caption{ANN Architecture. Such architecture is typically used by ML for image recognition}
    \label{fig:ANN_Architecture}
\end{figure}
If the architecture shown in Fig. \ref{fig:ANN_Architecture} is compared to that of Fig. \ref{fig:Chimera}, which shows a single cell with eight interconnected qubits within D-Wave's Chimera QPU in their quantum annealers, the similarity between the two architectures is immediately obvious. It may readily be seen how QPU architecture may be leveraged directly to implement ML based on neural network algorithms.

\section{Applications of Quantum Computing in P5 Medicine}
\label{sec:QuantumCure}
The following subsections describe a summary of current research where QC promised advantages over classical methods within the realm of P5 medicine.
%\subsection{Enhanced Data Analytics Capabilities}
%Finding Correlations
\subsection{Genome Sequencing}
Deoxyribonucleic acid (DNA), offers exceptionally dense information store \cite{heinis2023survey}. One gram of DNA can store 455 million terabytes (information density of 455 EB/g) \cite{church2012next}. This makes it possible, for a whole genome to fit into a human cell. QC can offer significant advantages over classical computing by enabling the processing of vast amount of genetic data more efficiently and accurately \cite{chow2024quantum}. 
QC can handle complex calculations and large datasets at unprecedented speeds, which is crucial for analyzing the massive amounts of data generated in genome sequencing \cite{nalkecz2022algorithm}. Additionally, QC can potentially reduce errors in sequencing by providing more precise calculations, leading to better identification of genetic variations and mutations \cite{chakraborty2023quantum}. The integration of QC with machine learning techniques can enhance predictive models and improve the interpretation of genetic data, leading to more personalized and effective treatments. As an example, quantum tensor decomposition has been successfully used to analyze high-dimensional, large-scale multi-omics data \cite{burch2025towards}. 

 Research even suggests, that DNA already functions as a topological quantum computer \cite{pitkanen2011dna}. Therefore, QC can help us understand DNAs structure and natural processes, and makes it a powerful and efficient system for processing genetic information \cite{riera2024dna}. DNA’s nitrogenous bases (Adenine, Thymine, Guanine, and Cytosine) form quantum states that can be seen as qubits and the hydrogen bonds between base pairs (A-T and G-C) can be seen as a form of quantum entanglement likened to Josephson Junctions. In QC, entangled qubits are linked such that the state of one qubit directly affects the state of another, no matter the distance between them. Similarly, the pairing of bases in DNA ensures that the state of one base is directly related to its pair. Additionally, the human body's ability to process DNA information is compared to quantum parallelism, where multiple computations occur simultaneously. Just as a QC can perform many calculations at once, genetic instructions from DNA can be transcribed simultaneously. DNA's ability to facilitate parallel processing allows it to be harnessed for uses which cannot be achieved using classical systems. This parallel processing capability could explain how DNA manages the vast amount of information required for cellular functions, replication, and repair so efficiently. 
 
\subsection{Drug Discovery}
With Eroom effect (Moore’s Law backwards), the pharmacutical industry faces a the critical challenge, that drug discovery is becomming exponentially more expensive over time \cite{hall2018paradox}. Designing new pharmaceuticals is a complex process which is already supported by computers, so called computer-aided drug design (CADD). The process of designing and optimizing drug molecules involves exploring vast chemical spaces to identify compounds with desired properties. Current, classical CADD tools cannot fully capture the complexity and nuances of real-world biological systems, leading to limitations in predicting drug efficacy and safety. Additionally, classical CADD is computationally expensive which limits its scalability to handle large datasets \cite{pei2024computer}. Although AlphaFold, developed by DeepMind, made significant advances in the prediction of 3D structure of proteins from their amino acid sequences, it is limited by classical computational power \cite{desai2024review}. While AlphaFold excels at predicting static structures, it does not account for the dynamic nature of proteins in a biological context. AlphaFold's predictions are less accurate for highly variable sequences, such as those found in immune system molecules like antibodies. Additionally, it is not sensitive to point mutations that change a single residue and also faces challenges with orphan proteins, which lack close relatives for comparison. 

\subsubsection{Molecular Modeling and Simulations}
QC can surpass limitations described above by efficiently searching through chemical  spaces and predicting the properties of new molecules. This can significantly accelerate the identification of promising drug candidates and optimize their efficacy and safety profiles. QC also has the potential to excel in simulating molecular interactions with high precision. Traditional computers struggle with the exponential complexity of these simulations, often relying on approximations. QC, however, can model quantum mechanical interactions directly, providing more accurate predictions of molecular behavior. It can simulate intrinsically disordered regions and protein dynamics more accurately to model complex molecular interactions, e.g., by improving the prediction of protein-protein interactions and post-translational modifications by providing more precise calculations of binding energies and conformational changes \cite{au2023np}. Additionally, quantum algorithms can handle highly variable sequences and point mutations more effectively, offering deeper insights into the structural impacts of these variations. This capabilities are crucial for understanding how drug candidates interact with biological targets, potentially leading to the discovery of more effective drugs.  Algorithms such as the Variational Quantum Eigensolver (VQE) and Quantum Phase Estimation (QPE) are used to solve the Schrödinger equation for complex molecules, providing insights into their electronic structures \cite{blunt2022perspective,hermann2020deep}. In quantum chemistry, QC can handle the many-body problem more efficiently, allowing for precise calculations of molecular properties like energy levels, bond lengths, and reaction pathways \cite{cheng2020application}.
QML algorithms can be used to predict molecular properties and optimize drug candidates by learning from quantum data. Quantum annealing can solve optimization problems by finding the global minimum of a function, which is useful in identifying the most promising molecular structures optimizing strong interactions with active sites with low interactions otherwise and high molecular stability.

QC can also improve the modeling of pharmacokinetics (how drugs are absorbed, distributed, metabolized, and excreted) and pharmacodynamics (the effects of drugs on the body). By simulating those two processes with greater accuracy, QC can help predict the behavior of drug candidates in the human body, leading to better-informed decisions in drug development. Quantum simulations can model complex biochemical interactions involved in pharmacokinetics and pharmacodynamics with high precision, while quantum-enhanced ML algorithms can analyze large datasets from clinical trials and predict drug behavior more accurately.

\subsubsection{Protein Folding}
Understanding protein folding is essential for drug discovery, as misfolded proteins are often implicated in diseases. QC can simulate the folding process more accurately than classical methods, providing insights into the structure and function of proteins. This can aid in the design of drugs that target specific protein conformations. Quantum Monte Carlo simulation methods can be used to simulate the thermodynamics of protein folding, providing detailed insights into the folding pathways and energy landscapes. Also, Quantum Annealing can be applied to solve the protein folding problem by finding the lowest energy conformation of a protein.

\subsubsection{Virtual Screening}
Virtual screening involves evaluating large libraries of compounds to identify potential drug candidates. QC can perform these screenings more efficiently by leveraging quantum algorithms to process and analyze data at a much faster rate than classical computers \cite{mensa2023quantum}. This can reduce the time and cost associated with early-stage drug discovery. Grover's algorithm can be used to search unsorted databases exponentially faster than classical algorithms, making it ideal for virtual screening. QC can also improve the accuracy of molecular docking simulations, which predict how small molecules, such as drug candidates, bind to a receptor.

\subsection{Predictive Modeling of Diseases and Treatment Outcomes}
The above-mentioned techniques can not only be used for drug discovery but also for disease prediction and data analytics. By rapidly analyzing extensive genomic, proteomic, and neuroimaging datasets, quantum algorithms can identify biomarkers associated with specific diseases, facilitating early detection and monitoring \cite{brainsci14010093}. Health data can be encoded in qubits, utilizing quantum entanglement and parallelism to process multiple data combinations simultaneously, allowing significantly improved computational speed and accuracy in uncovering higher-order correlations and early disease prediction compared to traditional methods \cite{mei6framework}. Further research has shown this, harnessing QML to outperform traditional ML for predicting heart diseases \cite{babu2024revolutionizing}. 

As mentioned above, QC can also simulate complex molecular interactions, providing insights into the pathogenesis of diseases like Alzheimer’s and Parkinson’s \cite{swarna2021parkinson}. Additionally, QC can enhance predictive modeling strategies by simultaneously analyzing genetic, behavioral, clinical, and environmental data, thereby identifying risk factors and enabling early intervention for mental health disorders \cite{brainsci14010093}. This capability to analyze vast amounts of patient data can be used to predict individual responses to therapies. As an example, quantum deep reinforcement learning algorithms have been explored for optimal decision-making in adaptive radiotherapy, potentially leading to more effective and tailored treatment strategies \cite{niraula2021quantum}.

QML can also be utilized to predict treatment outcomes by modeling tumor dynamics and patient responses. For example, hybrid quantum-classical neural architectures have been proposed to quantify tumor burden concerning treatment effects, predicting therapy responses and facilitating personalized medicine approaches \cite{nguyen2023translationalquantummachineintelligence}.

\subsection{Clinical Trial Optimization}
As emphasized by \cite{doga2024can} as well as \cite{kumar2024recent}, QC can address current challenges of clinical trials such as site selection, and cohort identification. Trial simulations and optimization has the potential to drastically reduce cost by allowing more efficient planning of clinical trials \cite{martin2017much}. Key technical aspects include the use of QML and quantum optimization algorithms to enhance various stages of clinical trials. For instance, quantum differential solvers can improve the accuracy of physiology-based pharmacokinetics and pharmacodynamics (PBPK/PD) models, which are crucial for predicting drug effects across different populations. Additionally, variational quantum algorithms (VQAs) and the quantum approximate optimization algorithm (QAOA) are highlighted for their ability to optimize trial site selection and cohort identification by efficiently exploring high-dimensional parameter spaces. Additionally, quantum generative models such as quantum Boltzmann machines, can be used to create synthetic patient data, which then can be used to simulate clinical trials and improve cohort identification \cite{sinno2025implementinglargequantumboltzmann}.

%%%%%%%%%%%%%%%%%%%%%%%%%%%%%%%%%%%%%%%%%%%%%%%%%%%%%%%%%%%%%%%
%
% Discussing challenges and opportunities of applying QC/QML in healthcare
% Showing how formal methods and validation can be used to enhance QC/QML
% Giving an outlook of the future of QC/QML in healthcare
%
%%%%%%%%%%%%%%%%%%%%%%%%%%%%%%%%%%%%%%%%%%%%%%%%%%%%%%%%%%%%%%%
\section{Discussion}
\label{sec:Discussion}
QC presents both exciting opportunities and significant challenges for P5 medicine, which emphasizes predictive, preventive, personalized, participatory, and psychocognitive healthcare. One of the primary technical challenges is scalability. Current quantum systems are limited by the number of qubits they can manage effectively, and maintaining coherence among these qubits is difficult due to environmental noise and decoherence. This necessitates robust error correction methods, which are still under development. Another challenge is the development of algorithms that can leverage quantum advantages for medical applications. While theoretical algorithms like Shor’s and Grover’s have shown promise, practical implementations for complex medical problems are still in their infancy. Integration with classical systems is also a significant hurdle. QC needs to work seamlessly with classical systems to be practical for medical applications, requiring efficient data transfer and hybrid computing models. Additionally, ensuring the privacy and security of sensitive medical data in quantum environments is crucial. Quantum cryptography offers potential solutions, but widespread implementation and standardization are still needed.

Despite these challenges, several QC techniques are showing promise and are ready for initial applications in P5 medicine. QML has shown competitive results with classical machine learning in various medical applications, such as drug discovery, medical imaging, and personalized treatment plans. Near-term quantum algorithms are being trained with diverse clinical datasets, demonstrating potential in predictive analytics and diagnostics. Quantum simulations are being used to model complex biological systems and molecular interactions, which can accelerate drug discovery and development. These simulations can provide insights that are difficult to achieve with classical computing. Quantum key distribution (QKD) is being explored to enhance the security of medical data, potentially providing unbreakable encryption and ensuring the confidentiality of patient information.

However, certain areas of QC still require significant advancements before they can be fully integrated into P5 medicine. More robust error correction methods are needed to make quantum computations reliable for medical applications. Current techniques are not yet sufficient to handle the error rates in large-scale quantum systems. Advances in quantum hardware, including the development of more stable qubits and better quantum processors, are essential. Innovations in materials science and quantum chip design will play a crucial role. Establishing standardized protocols for QC in healthcare is necessary, including developing guidelines for data handling, algorithm validation, and integration with existing medical systems. It also must to be taken account that quantum supremacy,  the ability to reduce the complexity of computationally expensive to solve problems, has only been shown for a limited number of tasks. QC is limited by the destructive nature of qubit measurement. This makes debugging or testing as we know it from classical programming challenging. Additionally, quantum hardware is error prone and algorithms complex. Formal methods, mathematically-based techniques for specifying, developing, and verifying software and hardware systems, can significantly enhance the reliability and correctness of QC applications in P5 medicine \cite{chareton2021formal} by providing a rigorous mathematical framework, which helps developers specify, develop, and verify systems with high precision. For instance, in genomic data analysis, researchers can use formal specification languages to precisely define the behavior and properties of a quantum algorithm designed to identify genetic markers associated with diseases. Model checking tools can then systematically explore all possible states of the algorithm to ensure it behaves correctly under all conditions, while theorem proving techniques provide mathematical proof that the algorithm meets its specified properties, ensuring accuracy and reliability \cite{lewis2023formal}. Additionally, formal optimization techniques can enhance the efficiency and performance of the quantum algorithm by reducing resource usage, such as the number of qubits and gate operations. \cite{quist2024advancing} shows based on a tutorial how two formal methods (\#SAT and decision diagrams (DD)) can be applied to quantum circuits. But the relationship is not unidirectional. QC can also enable formal methods such as verification tools, which are computationally expensive, to be applied to more complex tasks. Those applications of formal methods pose an area of future research. Apart from the technical issues and research areas such as algorithm design, also the ethical implications of QC in P5 medicine, particularly concerning data ownership, consent, and the potential for algorithmic bias should be explored further. %cite smartEHR paper

Nevertheless, the future of QC in P5 medicine is promising. As QC continues to evolve, we can expect several key developments. QC will enable more accurate predictive models for disease progression and treatment outcomes, leading to more personalized and effective healthcare. Quantum simulations will significantly speed up the drug discovery process, allowing for the rapid identification of new therapeutic compounds. Quantum-enhanced imaging and diagnostic tools will provide earlier and more accurate detection of diseases, improving patient outcomes. Quantum cryptography will ensure the highest levels of security for medical data, protecting patient privacy and fostering trust in digital health solutions. While there are substantial technical challenges to overcome, the advancements in QC hold great promise for P5 medicine. Continued research and collaboration between computer, data, and quantum scientists, as well as medical professionals, will be crucial in unlocking the full potential of this transformative technology.

\section{Conclusion}
\label{sec:Conclusion}
QC and QML hold transformative potential for advancing P5 medicine by offering unprecedented computational power and accuracy. These technologies can revolutionize diagnostics, personalized treatments, and drug discovery. However, challenges remain, including technical limitations, high costs and complexity, and the need for interdisciplinary collaboration. Formal methods can enhance the reliability and correctness of QC applications, ensuring accuracy and efficiency. Future research should focus on overcoming these challenges, implementing the suggested concepts, and exploring ethical implications. By addressing these issues, QC can significantly improve patient outcomes, contribute to the evolution of P5 medicine, and ultimately lead to more healthy life years for patients.
%\newpage
%\begin{credits}
%\subsubsection{\ackname}% A bold run-in heading in small font size at the end of the paper is used for general acknowledgments, for example: This study was funded by X (grant number Y).
%\subsubsection{\discintname}
%The presented research has been funded by Unisys' Enterprise Computing Solutions Unit.
%\end{credits}

\bibliographystyle{splncs04}
\bibliography{QuantumCure}
\end{document}